\documentclass[12pt]{iopart}

\usepackage{iopams,bm}
\usepackage{cite}

\newcommand{\hook}
{\raisebox{-0.35ex}{\makebox[0.6em][r]{\scriptsize $-$}}
\hspace{-0.15em}\raisebox{0.25ex}{\makebox[0.4em][l]{\tiny $|$}}}

\makeatletter

\@addtoreset{equation}{section}
\makeatother

\begin{document}

\begin{flushright}
DAMTP-2012-83, OCU-PHYS-377
\end{flushright}

\title[]{Killing--Yano symmetry of Kaluza--Klein black holes in five dimensions}
\author{Tsuyoshi Houri$^{1),2)}$ and Kei Yamamoto$^{1),3)}$}
\address{1) DAMTP, Centre for Mathematical Sciences, University of Cambridge,
 Wilberforce Road, Cambridge CB3 0WA, United Kingdom \\
 2) Osaka City University Advanced Mathematical Institute (OCAMI),
 3-3-138 Sugimoto, Sumiyoshi, Osaka 558-8585, Japan \\
 3) Institute of Theoretical Astrophysics, University of Oslo,
 P.O. Box 1029 Blindern, N-0315 Oslo, Norway}
 \eads{\mailto{T.Houri@damtp.cam.ac.uk; houri@sci.osaka-cu.ac.jp}, \mailto{K.Yamamoto@damtp.cam.ac.uk}}
 
\date{\today}

\begin{abstract}
 Using a generalised Killing--Yano equation in the presence of torsion,
 spacetime metrics admitting a rank-2 generalised Killing--Yano tensor are investigated
 in five dimensions under the assumption that its eigenvector associated with the zero
 eigenvalue is a Killing vector field.
 It is shown that such metrics are classified into three types
 and the corresponding local expressions are given explicitly.
 It is also shown that they cover some classes of charged, rotating Kaluza--Klein black hole solutions
 of minimal supergravity and abelian heterotic supergravity.
\end{abstract}


\maketitle

\section{Introduction}
Killing--Yano tensors which were introduced by K.\ Yano \cite{Yano:1952}
as a generalisation of Killing vectors to higher-rank antisymmetric tensors,
have attracted the interests of many authors in the study of black hole physics
as their presence endows black hole spacetimes with remarkable mathematical properties.
For instance, B.\ Carter \cite{Carter:1968} studied a certain class of metrics
including the Kerr metric as well as Kerr--Newman metric,
\begin{equation}
\eqalign{
 ds^2 =& \frac{r^2+p^2}{{\cal Q}(r)}dr^2 + \frac{r^2+p^2}{{\cal P}(p)}dp^2  \\
       & -\frac{{\cal Q}(r)}{r^2+p^2}(d\tau + p^2 d\sigma)^2 
           + \frac{{\cal P}(p)}{r^2+p^2}(d\tau - r^2 d\sigma)^2 ~, }\label{Carter metric}
\end{equation}
which is called Carter's class. 
He demonstrated that for all the metrics of this class,
both the Hamilton--Jacobi and Klein--Gordon equations 
can be solved by separation of variables.
These separability structures are known to be deeply related to 
the existence of rank-2 Killing--Yano tensors \cite{Carter:1977,Carter:1979}.
Moreover, W.\ Dietz and R.\ Rudiger \cite{Dietz:1982} have shown that
any four-dimensional metric admitting a rank-2 Killing--Yano tensor can be always written 
in the form of Carter's class.
Namely, in four dimensions spacetime metrics admitting a rank-2 Killing--Yano tensor
involve many rotating black hole solutions of Einstein's equations, e.g., Kerr and Kerr--Newman metrics,
and due to the Killing--Yano symmetry,
some test field equations such as Hamilton--Jacobi and Klein--Gordon equations
are solvable by separation of variables.
One of our tasks remaining in this direction
is clarifying what happens in higher dimensions.

The investigation of Killing--Yano symmetry
in higher-dimensional black hole spacetimes has just started in the last decade,
e.g., see reviews \cite{Frolov:2008,Yasui:2011,Santillan:2011}.
It is known that in higher-dimensional vacuum solutions 
describing rotating black holes with spherical horizon topology
\cite{Myers:1986,Hawking:1999,Gibbons:2004,Chen:2006},
rank-2 closed conformal Killing--Yano tensors explain
the separability of the Hamilton--Jacobi and Klein--Gordon equations on those spacetimes.
Here, conformal Killing--Yano (CKY) tensors are antisymmetric tensors
which were introduced by S.\ Tachibana \cite{Tachibana:1969} and T.\ Kashiwada \cite{Kashiwada:1968}
as a generalisation of conformal Killing vectors.
However, CKY tensors are no longer useful to explain separability structure
in higher-dimensional charged, rotating black hole spacetimes.
Accordingly, a further generalisation of CKY tensors
was introduced by the authors of \cite{Kubiznak:2009,Kubiznak:2011}.
The generalised CKY tensors can be thought of as 
CKY tensors on spacetimes with a skew-symmetric torsion $\bm{T}$.
The torsion is usually (not necessarily) identified with matter fluxes appearing in the theories.
For instance, the five-dimensional gauged minimal supergravity black hole \cite{Chong:2005}
admits a rank-2 generalised CKY tensor
when the torsion is identified with the Hodge dual of the Maxwell filed,
$\bm{T}=*\bm{F}/\sqrt{3}$.
It was also shown that the abelian heterotic supergravity black holes
as well as their generalisation to higher dimensions \cite{Sen:1992,Cvetic:1996,Chow:2010}
admit a rank-2 generalised CKY tensor.
In this case, the torsion is identified with the 3-form field strength, $\bm{T}=\bm{H}$.

Compared to the asymptotically flat black holes, the separability structure of black strings
or Kaluza--Klein black holes has not been studied very well. Since the inheritance of separability
by the uplift which is obvious for vacuum solutions is also expected in the presence of
flux along the extra dimension, we are lead to the investigation of the generalised Killing--Yano
symmetry in those spacetimes.
In this paper, we thus elaborate the relationship between Killing--Yano symmetry and separability
of the Hamilton--Jacobi and Klein--Gordon equations
in charged, rotating Kaluza--Klein black hole spacetimes,
especially in the five-dimensional minimal supergravity
and abelian heterotic supergravity.
In fact, it is shown that
Killing--Yano symmetry of the Kaluza--Klein black holes
are described by rank-2 generalised Killing--Yano tensors.
While Killing--Yano and generalised Killing--Yano tensors have been discussed
in the relation to separation of Dirac equations as well
\cite{McLenaghan:1979,Kamran:1984,Benn:1997,Benn:2004,Houri:2010b,Cariglia:2011,Cariglia:2011b,Carignano:2011},
we will not discuss it in this article.

The presence of Killing--Yano symmetry itself is a strong enough restriction so that
one can obtain explicit expressions of the metrics before imposing the dynamical equations.
For example, some geometrical properties of spacetimes admitting
Killing--Yano tensors were discussed in four dimensions \cite{Dietz:1982,Dietz:1981,Spindel:1979}.
The most general metric admitting a rank-2 closed CKY tensor
was obtained in arbitrary dimension \cite{Krtous:2008,Houri:2009}.
Even when it is not possible to write down the most general metric,
this approach enables us to understand separability structure 
of various black hole spacetimes in a wider and unified framework \cite{Houri:2010,Houri:2012}.
By making a suitable simplification, we indeed derive a class of five-dimensional metrics
admitting a rank-2 generalised Killing--Yano tensor, which include and generalise the known
examples of black strings and Kaluza--Klein black holes.

This paper is organised as follows: In Sec.\ 2,
we first attempt to classify spacetime metrics
admitting a rank-2 generalised Killing--Yano tensor in five dimensions
 under the assumption that there exists
a particular Killing vector field.
A large family of metrics is obtained.
We find that resulting metrics are classified into three types,
which we call type A, B and C, and
some local expressions of the metrics are given explicitly.
In Sec.\ 3, we consider the solution in the five-dimensional minimal supergravity,
which is obtained as an uplift of the Kerr spacetime,
and see that the metric falls into type A of the classified metrics.
It is shown that a rank-2 generalised Killing--Yano tensor is responsible for
separation of variables in the Hamilton--Jacobi and Klein--Gordon equations.
Using the type A metric classified in Sec.\ 2,
we construct the general solution in the five-dimensional minimal supergravity.
In Sec.\ 4, we review the charged, rotating black string solution discovered by Mahapatra \cite{Mahapatra:1994}
in heterotic supergravity.
We will see that the metric falls into type A again.
The separability of the Hamilton--Jacobi and Klein--Gordon equations
is also associated to a rank-2 generalised Killing--Yano tensor.
Adopting the generalised Killing--Yano symmetry,
we construct a class of charged, rotating Kaluza--Klein black hole solutions
in the theory. Sec.\ 5 is devoted to summary and discussion.

\section{Metrics admitting a rank-2 generalised Killing--Yano tensor in five dimensions}
In this section, we attempt to classify spacetime metrics
admitting a rank-2 generalised Killing--Yano tensor
in five dimensions. Although we are interested in Lorentzian manifolds,
for simplicity, the calculation in this section is carried out
in Euclidean signature $(+,+,\cdots,+)$. Keeping applications to Kaluza--Klein spacetimes in mind,
we assume that there exists a particular Killing vector field that is
an eigenvector of the generalised Killing--Yano tensor with zero eigenvalue, 
leaving more general investigations for a future work.
In Sec.\ 2.1, we begin with reviewing the basics of the rank-2 generalised Killing--Yano tensors.
Introducing canonical frames associated with such tensors,
we derive the general forms of connection 1-forms in terms of the canonical frame in Sec.\ 2.2.
The computation there is performed by exploiting the technique of \cite{Houri:2012}.
Furthermore, we proceed to restrict the forms of the connection 1-forms
by imposing integrability conditions,
so that commutation relations among canonical basis vectors are obtained in Sec.\ 2.3.
Finally, solving the commutation relations in Sec.\ 2.4--2.6,
some local expressions of the metrics are given explicitly.
In the process of solving the commutation relations,
we find that resulting metrics are classified into three types,
which we call type A, B and C.

\subsection{Basics}
Let $(M,\bm{g})$ be a five-dimensional Riemannian manifold
and $\{\bm{e}_a\}$ be an orthonormal frame.
Throughout the article, Latin indices $a,b,\cdots$ range from 1 to 5.
Greek letters $\mu , \nu \cdots $ will later be used to denote two-dimensional
eigenspaces of non-zero eigenvalues of a 2-form.
The dual frame $\{\bm{e}^a\}$ satisfies $\bm{e}_a\hook\bm{e}^b = \delta_a{}^b$ 
where $\hook$ represents the inner product.
A $p$-form $\bm{k}$ is written as
\begin{equation}
 \bm{k} = \frac{1}{p!}\,k_{a_1\cdots a_p} \, \bm{e}^{a_1} \wedge \bm{e}^{a_p} ~,~~~
 k_{[a_1\cdots a_p]} = k_{a_1\cdots a_p} ~.
\end{equation}

For a 2-form $\bm{f}$, a rank-2 Killing--Yano tensor, 
introduced by \cite{Yano:1952}, is subject to the equation
\begin{equation}
 \nabla_a f_{bc} + \nabla_b f_{ca} = 0 ~, \label{KYeq}
\end{equation}
where $\nabla_a$ is the Levi-Civita connection.
In the presence of a skew-symmetric torsion, $T_{[abc]}=T_{abc}$,
a connection $\nabla^T_a$ is defined by
\begin{equation}
 \nabla^T_a X^b = \nabla_a X^b + \frac{1}{2} T_{ac}{}^b X^c ~.
\end{equation}
By replacing the connections $\nabla_a$ in (\ref{KYeq})
with $\nabla^T_a$, rank-2 generalised Killing--Yano (GKY) tensors are defined \cite{Kubiznak:2009} by
\begin{equation}
 \nabla^T_{a}f_{bc} + \nabla^T_{b}f_{ac} = 0 ~. \label{GKYEq}
\end{equation}

For a rank-$p$ GKY tensor, the Hodge dual gives 
a rank-$(D-p)$ generalised closed conformal Killing--Yano (GCCKY) tensor in $D$ dimensions \cite{Houri:2010}.
In $D=5$, the Hodge dual $\bm{h}=* \bm{f}$ of a rank-2 GKY tensor $\bm{f}$
is a rank-3 GCCKY tensor obeying
\begin{equation}
 \nabla^T_ah_{bcd} = g_{ab}\xi_{cd}+g_{ac}\xi_{db}+g_{ad}\xi_{bc} ~, \label{CCKY Eq Comp}
\end{equation}
where
\begin{equation}
 \xi_{ab} = \frac{1}{3}\nabla^{Tc}h_{cab} \label{Associated Comp}
\end{equation}
is called an associated 2-form of $\bm{h}$.
Eq.\ (\ref{CCKY Eq Comp}) implies that
\begin{equation}
 \nabla^T_{[a}h_{bcd]} = 0 ~,~~~ \nabla^T{}^b\xi_{ba} = 0 ~.
\end{equation}

From a rank-2 GKY tensor $\bm{f}$, one can construct a rank-2 Killing--St\"ackel tensor
\begin{equation}
 K_{ab} = f_{ac}f_b{}^c \label{KYtoKT} \ ,
\end{equation}
which is characterised by $\nabla _{(a}K_{bc)}=0$. In general, the existence
of a rank-2 Killing-St\"ackel tensor guarantees separation of the Hamilton--Jacobi 
equations for geodesics with the corresponding separation constant 
$\kappa^{(HJ)}$ given by
\begin{equation}
 \kappa^{(HJ)} = K_{ab}\Pi^a\Pi^b ~, \label{constantFromKS}
\end{equation}
where $\Pi^a$ is the canonical momentum associated with the geodesic.
When $K_{ab}$ is written in terms of a rank-2 Killing--Yano tensor as (\ref{KYtoKT}),
the separation constant of the Klein--Gordon equation $\kappa^{(KG)}$ appears
as an eigenvalue of a symmetry operator \cite{Carter:1968,Carter:1977,Carter:1979}
\begin{equation}
\hat{K} \equiv \nabla_a K^{ab} \nabla_b \label{symmetryOperator}
\end{equation}
that satisfies $[\hat{K},\Box]=0$ where $\Box$ is the scalar wave operator
$\Box\equiv g^{ab}\nabla_a\nabla_b$.
That is,
\begin{equation}
 \hat{K}\Psi = \kappa^{(KG)} \Psi ~.
\end{equation}
However, this is not always true for rank-2 GKY tensors.
In fact, we will see that separation of variables occurs
in a deformed Klein--Gordon equation in Sec.\ 4.

\subsection{General forms of the connection 1-forms}
Let us consider a rank-3 GCCKY tensor $\bm{h}$ in five dimensions,
which is equivalent to considering a rank-2 GKY tensor $\bm{f}$.
Then one can always find an orthonormal frame 
$\{\bm{e}^a\}=\{\bm{e}^\mu,\bm{e}^{\hat{\mu}}=\bm{e}^{2+\mu},\bm{e}^0=\bm{e}^5\}$, $\mu = 1,2$,
such that a metric $\bm{g}$ and a rank-3 GCCKY tensor $\bm{h}$ are
simultaneously written in the form
\begin{eqnarray}
 \bm{g} &=& \sum_{\mu=1}^2 (\bm{e}^\mu\otimes \bm{e}^\mu+\bm{e}^{\hat{\mu}}\otimes \bm{e}^{\hat{\mu}})
            +\bm{e}^0\otimes \bm{e}^0 ~, \\
 \bm{h} &=& \sum_{\mu=1}^2 x_\mu\,\bm{e}^\mu\wedge \bm{e}^{\hat{\mu}}\wedge \bm{e}^0 ~, \label{metand3form}
\end{eqnarray}
where $x_\mu$ are called the eigenvalues of $\bm{h}$.
The rank-3 GCCKY tensor is said to be non-degenerate 
if its eigenvalues $x_\mu$ are non-vanishing functions with $x_1\neq x_2$. 
Since there are still degrees of freedom under rotation in each 
$(e_\mu,e_{\hat{\mu}})$-plane,
the orthonormal frame is fixed completely by introducing a 1-form $\bm{\eta}$ as
\begin{equation}
  \bm{\eta} = -\bm{e}_0\hook \bm{\xi}
            = \sqrt{Q_1}\, \bm{e}^{\hat{1}}+\sqrt{Q_2}\, \bm{e}^{\hat{2}} ~,
\end{equation}
where $\bm{\xi}$ is the associated 2-form introduced in (\ref{Associated Comp})
and $Q_1$ and $Q_2$ are unknown functions.
This means that we used the remaining rotations so as to set $\xi_{1 0}=\xi_{20}=0$.
The fixed orthonormal frame is called a canonical frame.
With respect to the canonical frame, the rank-2 GKY tensor $\bm{f}$
and rank-2 Killing--St\"ackel tensor $\bm{K}$, given by (\ref{KYtoKT}), are written as
\begin{eqnarray}
 \bm{f} &=& x_2 \,\bm{e}^1\wedge \bm{e}^{\hat{1}} + x_1 \,\bm{e}^2\wedge\bm{e}^{\hat{2}} ~, \\
 \bm{K} &=& x_2^2 \,(\bm{e}^1\otimes \bm{e}^1+\bm{e}^{\hat{1}}\otimes \bm{e}^{\hat{1}})
            +x_1^2 \,(\bm{e}^2\otimes \bm{e}^2+\bm{e}^{\hat{2}}\otimes \bm{e}^{\hat{2}}) ~.
\end{eqnarray}

The GCCKY equation (\ref{CCKY Eq Comp}) can be thought of 
as relating components of connection 1-forms $\bm{\omega}^a{}_b$
to the eigenvalues $x_\mu$ of the GCCKY tensor $\bm{h}$ and their derivatives.
Exploiting the technique of \cite{Houri:2012} which was developed for rank-2 GCCKY tensors,
we now apply it to rank-3 GCCKY tensors in five dimensions.
Thus a similar calculation leads us to the following results:
\begin{eqnarray}
 \bm{\omega}{}^\mu{}_\nu
&=& -\frac{x_\nu\sqrt{Q_\nu}}{x_\mu^2-x_\nu^2}\,\bm{e}^\mu
   -\frac{x_\mu\sqrt{Q_\mu}}{x_\mu^2-x_\nu^2}\,\bm{e}^\nu
   -\frac{x_\nu\xi_{\mu\hat{\nu}}+x_\mu\xi_{\nu\hat{\mu}}}{x_\mu^2-x_\nu^2}\,\bm{e}^0 ~, \label{con1} \\
 \bm{\omega}{}^\mu{}_{\hat{\mu}}
&=& -\frac{1}{\sqrt{Q_\mu}}\frac{x_\mu Q_\nu}{x_\mu^2-x_\nu^2}\, \bm{e}^{\hat{\mu}} 
   +\frac{x_\mu\sqrt{Q_\nu}}{x_\mu^2-x_\nu^2}\, \bm{e}^{\hat{\nu}} \label{con2} \\
 && -\frac{\sqrt{Q_\nu}}{\sqrt{Q_\mu}}\frac{x_\nu\xi_{\mu\nu}
            +x_\mu\xi_{\hat{\mu}\hat{\nu}}}{x_\mu^2-x_\nu^2}\, \bm{e}^0
 +\sum_a\frac{\kappa_{a\mu}}{\sqrt{Q_\mu}}\,\bm{e}^a ~, \label{con3} \\
 \bm{\omega}{}^\mu{}_{\hat{\nu}}
&=& \frac{x_\mu\sqrt{Q_\nu}}{x_\mu^2-x_\nu^2}\, \bm{e}^{\hat{\mu}} 
   -\frac{x_\mu\sqrt{Q_\mu}}{x_\mu^2-x_\nu^2}\, \bm{e}^{\hat{\nu}}
   +\frac{x_\nu\xi_{\mu\nu}+x_\mu\xi_{\hat{\mu}\hat{\nu}}}{x_\mu^2-x_\nu^2}\, \bm{e}^0 ~, \label{con4} \\
 \bm{\omega}{}^{\hat{\mu}}{}_{\hat{\nu}}
&=& -\frac{x_\mu\sqrt{Q_\nu}}{x_\mu^2-x_\nu^2}\, \bm{e}^\mu
   -\frac{x_\nu\sqrt{Q_\mu}}{x_\mu^2-x_\nu^2}\,\bm{e}^\nu
   -\frac{x_\mu\xi_{\mu\hat{\nu}}+x_\nu\xi_{\nu\hat{\mu}}}{x_\mu^2-x_\nu^2}\, \bm{e}^0 ~, \label{con5} \\
 \bm{\omega}{}^\mu{}_0
&=& \frac{\xi_{\nu\hat{\nu}}}{x_\nu} \,\bm{e}^\mu
   -\frac{\xi_{\mu\hat{\nu}}}{x_\nu} \,\bm{e}^\nu
   +\frac{\xi_{\mu\nu}}{x_\nu} \,\bm{e}^{\hat{\nu}} ~, \label{con6}\\
 \bm{\omega}{}^{\hat{\mu}}{}_0
&=& -\frac{\xi_{\hat{\mu}\hat{\nu}}}{x_\nu} \,\bm{e}^\nu
   +\frac{\xi_{\nu\hat{\nu}}}{x_\nu} \,\bm{e}^{\hat{\mu}}
   -\frac{\xi_{\nu\hat{\mu}}}{x_\nu} \,\bm{e}^{\hat{\nu}} ~, \label{con7}
\end{eqnarray}
where the symbols $\kappa_{ab}$ are defined as
\begin{equation}
 \kappa_{ab} = \bm{e}_b \hook \nabla^T_{\bm{e}_a}\bm{\eta} ~.
\end{equation}

\subsection{Integrability conditions}
To obtain the information about second derivatives $\kappa_{ab}$,
we consider the integrability conditions of the Killing--Yano equation (\ref{CCKY Eq Comp}),
\begin{equation}
\eqalign{
& -R^T{}^f{}_{ade}h_{fbc}-R^T{}^f{}_{bde}h_{fca}-R^T{}^f{}_{cde}h_{fab}  \\
& = g_{ae}\nabla^T_d\xi_{bc}+g_{be}\nabla^T_d\xi_{ca}+g_{ce}\nabla^T_d\xi_{ab}
     - (d\leftrightarrow e)  \\
& ~~~ + T_{ade}\xi_{bc} + T_{bde}\xi_{ca} + T_{cde}\xi_{ab} ~,} \label{IC}
\end{equation}
where $R^T{}^a{}_{bcd}$ are components of the Riemann tensor with respect to $\nabla^T_a$ defined by
\begin{equation}
 (\nabla^T_a\nabla^T_b-\nabla^T_b \nabla^T_a+T_{ab}{}^d\nabla^T_d) Z_c \equiv -R^T{}^d{}_{cab} Z_d ~.
\end{equation}
The components do not satisfy the Bianchi identities $R^T{}^a{}_{[bcd]}= 0$, while
they have the following symmetries among their indices:
\begin{equation}
 R^T{}_{bacd}=-R^T{}_{abcd} ~,~~~ 
 R^T{}_{abdc}=-R^T{}_{abcd} ~.
\end{equation}

The general form of the integrability conditions is too complicated
to solve analytically since it contains many coupled, nonlinear partial differential equations.
Therefore, for simplicity, we impose an assumption that $\bm{e}_0$ is a Killing vector field.
This is motivated by the fact that the known example which admits a rank-3 
GCCKY tensor in the literature \cite{Houri:2012b} satisfies this assumption.
It leads us to vanishing of many components of the associated 2-form $\bm{\xi}$ 
and the torsion $\bm{T}$.
In fact, the Killing equation $\nabla_a(\bm{e}^0)_b+\nabla_b(\bm{e}^0)_a=0$ implies
\begin{equation}
 \xi_{\mu\nu} = \xi_{\mu\hat{\mu}} = \xi_{\mu\hat{\nu}} = \xi_{\hat{\mu}\hat{\nu}} = 0 ~,
\end{equation}
and hence it follows from the integrability conditions (\ref{IC}) that
\begin{eqnarray}
 T_{\mu\nu\hat{\nu}}=T_{\mu\nu 0} = T_{\mu\hat{\nu}0} = T_{\hat{\mu}\hat{\nu}0}=0 ~, \nonumber \\
 T_{\mu\hat{\mu}0} = -\frac{\kappa_{0\mu}}{\sqrt{Q_\mu}} ~,~~~
 T_{\mu\hat{\mu}\hat{\nu}} = -\frac{\kappa_{\hat{\nu}\mu}}{\sqrt{Q_\mu}} ~.
\end{eqnarray}
The commutators now simplify to
\begin{eqnarray}
\left[\bm{e}_\mu,\bm{e}_\nu \right] 
= -\frac{x_\nu\sqrt{Q_\nu}}{x_\mu^2-x_\nu^2}\,\bm{e}_\mu
   -\frac{x_\mu\sqrt{Q_\mu}}{x_\mu^2-x_\nu^2}\,\bm{e}_\nu ~, \nonumber \\
\left[\bm{e}_\mu,\bm{e}_{\hat{\mu}}\right]
= K_\mu\,\bm{e}_\mu + L_\mu\,\bm{e}_{\hat{\mu}}
   + M_\mu\,\bm{e}_{\hat{\nu}} - T_{\mu\hat{\mu}0} \,\bm{e}_0 ~, \nonumber \\
\left[\bm{e}_\mu,\bm{e}_{\hat{\nu}}\right]
= -\frac{x_\mu\sqrt{Q_\mu}}{x_\mu^2-x_\nu^2}\,\bm{e}_{\hat{\nu}} ~, \label{c4-6}\\
\left[\bm{e}_{\hat{\mu}},\bm{e}_{\hat{\nu}}\right]
=0 ~,~~~ \left[\bm{e}_\mu,\bm{e}_0\right]=0 ~,~~~
\left[\bm{e}_{\hat{\mu}},\bm{e}_0\right] = 0 ~, \nonumber 
\end{eqnarray}
where
\begin{equation}
\eqalign{
  K_\mu = \frac{\kappa_{\mu\mu}}{\sqrt{Q_\mu}} ~,~~~
  L_\mu = -\frac{1}{\sqrt{Q_\mu}}\Big(\frac{x_\mu Q_\nu}{x_\mu^2-x_\nu^2}-\kappa_{\hat{\mu}\mu}\Big) ~,  \\
  M_\mu = \frac{2x_\mu\sqrt{Q_\nu}}{x_\mu^2-x_\nu^2}-T_{\mu\hat{\mu}\hat{\nu}} ~.}
\end{equation}

The classification problem under the assumption that $\bm{e}_0$ is a Killing vector field
reduces to solving the commutators (\ref{c4-6}).
The integrability conditions, e.g., the Jacobi identities, which have not been satisfied yet,
give rise to two algebraic equations
\begin{equation}
M_1K_2 = 0 ~,~~~ M_2K_1 = 0 ~, \label{AE}
\end{equation}
and a system of coupled, nonlinear partial differential equations
\begin{equation}
\eqalign{
& \bm{e}_{\nu } \left( K_{\mu } \right) = \frac{x_{\nu }\sqrt{Q_{\nu }}}{x_{\mu }^2 -x_{\nu }^2} K_{\mu }~ , \\
& \bm{e}_{\nu } \left( L_{\mu } \right) = \frac{x_{\nu } \sqrt{Q_{\nu }}}{x_{\mu }^2 -x_{\nu }^2} L_{\mu } - M_{\mu } M_{\nu } - \frac{2x_{\mu }x_{\nu } \sqrt{Q_{\mu }}\sqrt{Q_{\nu }}}{(x_{\mu }^2 -x_{\nu }^2)^2}~ , \\
& \bm{e}_{\nu } \left( M_{\mu } \right) = \left( \frac{2x_{\nu }\sqrt{Q_{\nu }}}{x_{\mu }^2 -x_{\nu }^2} -L_{\nu } \right) M_{\mu } ~, \\
& \bm{e}_{\nu }\left( T_{\mu \hat{\mu }0} \right) = \frac{2x_{\nu }\sqrt{Q_{\nu }}}{x_{\mu }^2 -x_{\nu }^2} T_{\mu \hat{\mu }0} - M_{\mu } T_{\nu \hat{\nu }0} ~, }
\end{equation}
and
\begin{equation}
\eqalign{
& \bm{e}_{\hat{\nu }} \left( K_{\mu }\right) = 0 ~, ~~~ \bm{e}_{\hat{\nu }} \left( L_{\mu } \right) = 0 ~, ~~~ \bm{e}_{\hat{\nu }} \left( M_{\mu } \right) = 0 ~,~~~ \bm{e}_{\hat{\nu }} \left( T_{\mu \hat{\mu }0} \right) = 0 ~,  \\
& \bm{e}_{0} \left( K_{\mu } \right) = 0 ~, ~~~ \bm{e}_0 \left( L_{\mu } \right) = 0 ~,~~~ \bm{e}_0 \left( M_{\mu } \right) = 0 ~,~~~  \bm{e}_0 \left( T_{\mu \hat{\mu } 0} \right) = 0 ~.}
\end{equation}
There exist orthonormal frames satisfying (\ref{c4-6}) at least
if its integrability conditions hold.
Namely, the algebraic equations (\ref{AE}) imply that
there are three types of the solutions:
A) $K_1=K_2=0$, B) $M_1=M_2=0$ and C) $K_1=M_1=0$ or $K_2=M_2=0$.
For each type, we are able to find large families 
of solutions of the remaining partial differential equations.

\subsection{Type A metric}

\subsection*{2.4.1. type A$_\lambda$}
Firstly, we consider $K_1=K_2=0$ case.
In this case, we find the canonical frame $\{\bm{e}_a\}$ satisfying
the commutators (\ref{c4-6}) as
\begin{equation}
\eqalign{
& \bm{e}_\mu = \sqrt{\frac{X_\mu}{x_\mu^2-x_\nu^2}} \frac{\partial}{\partial x_\mu} ~,~~~ 
  \bm{e}_0 = \frac{\partial}{\partial \psi} ~,  \\
& \bm{e}_{\hat{\mu}} = \frac{1}{f_\mu \sqrt{(x_\mu^2-x_\nu^2)X_\mu}}
                     \Big((x_\mu^2+N_\mu) \frac{\partial}{\partial \tau}- \frac{\partial}{\partial \sigma}
                     -\lambda N_\mu\frac{\partial}{\partial \psi}\Big) ~,} \label{eq2-35}
\end{equation}
where $X_\mu$, $N_\mu$ and $f_\mu$ are unknown functions of one variable $x_\mu$
and $\lambda$ is an arbitrary constant.
Rewriting $(x_1,x_2)$ by $(x,y)$,
the corresponding metric is written by
\begin{equation}
\fl \eqalign{
 \bm{g}\ =\ & \frac{x^2-y^2}{X(x)}dx^2+\frac{y^2-x^2}{Y(y)}dy^2
            + (d\psi + \lambda \bm{W}_1 )^2  \\
         &  + \frac{f_1(x)^2 X(x)}{x^2-y^2} (d\tau + y^2 d\sigma - \bm{W}_1)^2
           + \frac{f_2(y)^2 Y(y)}{y^2-x^2} (d\tau + x^2 d\sigma - \bm{W}_1)^2 ~,} \label{metAI}
\end{equation}
where 
\begin{equation}
\eqalign{
& \bm{W}_1 = \frac{N_1(x)}{(x^2-y^2)\Phi}(d\tau + y^2 d\sigma)
           + \frac{N_2(y)}{(y^2-x^2)\Phi}(d\tau + x^2 d\sigma) ~,  \\
& \Phi = 1 + \frac{N_1(x)}{x^2-y^2} + \frac{N_2(y)}{y^2-x^2} ~.}
\end{equation}
The metric contains 6 unknown functions of one variable,
$X(x)$, $Y(y)$, $N_1(x)$, $N_2(y)$, $f_1(x)$ and $f_2(y)$.
Since the metric components are independent of the coordinates $\tau$, $\sigma$ and $\psi$,
$\partial/\partial \tau$, $\partial/\partial \sigma$ and $\partial/\partial \psi$ 
are three Killing vector fields.
Then the torsion is given by
\begin{equation}
\fl \eqalign{
 \bm{T} \ = \ &  \Bigg[\frac{2x}{x^2-y^2}
           -\frac{f_2(y)}{f_1(x)}\Big(\frac{2x}{x^2-y^2}
            +\frac{\partial \ln \Phi}{\partial x} \Big)\Bigg] \sqrt{\frac{Y(y)}{y^2-x^2}}
           \,\bm{e}^1\wedge\bm{e}^{\hat{1}}\wedge\bm{e}^{\hat{2}}  \\
         & +\Bigg[\frac{2y}{y^2-x^2}
           -\frac{f_1(x)}{f_2(y)}\Big(\frac{2y}{y^2-x^2}
            +\frac{\partial \ln \Phi}{\partial y} \Big)\Bigg] \sqrt{\frac{X(x)}{x^2-y^2}}
           \,\bm{e}^2\wedge\bm{e}^{\hat{2}}\wedge\bm{e}^{\hat{1}}  \\
         & + \frac{\lambda}{f_1(x)} \frac{\partial \ln \Phi}{\partial x}
           \,\bm{e}^1\wedge\bm{e}^{\hat{1}}\wedge\bm{e}^0
           + \frac{\lambda}{f_2(y)} \frac{\partial \ln \Phi}{\partial y}
           \,\bm{e}^2\wedge\bm{e}^{\hat{2}}\wedge\bm{e}^0 ~.} \label{eq:torsionAI}
\end{equation}
This type of metrics includes
the charged, rotating black strings discovered by S.\ Mahapatra \cite{Mahapatra:1994}
and Kaluza--Klein black holes in abelian heterotic supergravity, 
as we shall see in Sec. 4.

\subsection*{2.4.2. type A$_\infty$}
An interesting special case arises in the limit $\lambda \to \infty$,
keeping fixed the product $\lambda N_\mu$.
To be explicit, if one scales $\lambda$ and $N_\mu$ in (\ref{eq2-35}) as follows,
\begin{equation}
 \lambda \to \lambda/\epsilon \,, \qquad N_\mu \to \epsilon N_\mu \,,
\end{equation}
and then takes the scaling limit $\epsilon \to 0$,
one obtains precisely
\begin{equation}
\eqalign{
& \bm{e}_\mu = \sqrt{\frac{X_\mu}{x_\mu^2-x_\nu^2}} \frac{\partial}{\partial x_\mu} ~,~~~ 
  \bm{e}_0 = \frac{\partial}{\partial \psi} ~,  \\
& \bm{e}_{\hat{\mu}} = \frac{1}{f_\mu \sqrt{(x_\mu^2-x_\nu^2)X_\mu}}
                     \Big(x_\mu^2 \frac{\partial}{\partial \tau}- \frac{\partial}{\partial \sigma}
                     -N_\mu \frac{\partial}{\partial \psi}\Big) ~,}
\end{equation}
where $X_\mu$, $N_\mu$ and $f_\mu$ are functions of single variable $x_\mu$ again.
Thus the corresponding metric is given by
\begin{equation}
\eqalign{
 \bm{g}\ =\ & \frac{x^2-y^2}{X(x)}dx^2+\frac{y^2-x^2}{Y(y)}dy^2
            + (d\psi + \bm{W}_2 )^2  \\
         &  + \frac{f_1(x)^2 X(x)}{x^2-y^2} (d\tau + y^2 d\sigma)^2
            + \frac{f_2(y)^2 Y(y)}{y^2-x^2} (d\tau + x^2 d\sigma)^2 ~,} \label{metAII}
\end{equation}
where
\begin{equation}
 \bm{W}_2 = \frac{N_1(x)}{x^2-y^2}(d\tau + y^2 d\sigma)
           + \frac{N_2(y)}{y^2-x^2}(d\tau + x^2 d\sigma) ~.
\end{equation}
The metric contains 6 unknown functions of one variable,
$X(x)$, $Y(y)$, $N_1(x)$, $N_2(y)$, $f_1(x)$ and $f_2(y)$
and has three Killing vector fields
$\partial/\partial \tau$, $\partial/\partial \sigma$ and $\partial/\partial \psi$.
The torsion is given by
\begin{equation}
\eqalign{
 \bm{T}\ =\ & \left(1-\frac{f_2(y)}{f_1(x)}\right)\frac{2x}{x^2-y^2}\sqrt{\frac{Y(y)}{y^2-x^2}}
           \,\bm{e}^1\wedge\bm{e}^{\hat{1}}\wedge\bm{e}^{\hat{2}}  \\
         & +\left(1-\frac{f_1(x)}{f_2(y)}\right)\frac{2y}{y^2-x^2}\sqrt{\frac{X(x)}{x^2-y^2}}
           \,\bm{e}^2\wedge\bm{e}^{\hat{2}}\wedge\bm{e}^{\hat{1}}  \\
         & + \frac{1}{f_1(x)} \frac{\partial  \Phi}{\partial x}
           \,\bm{e}^1\wedge\bm{e}^{\hat{1}}\wedge\bm{e}^0
           + \frac{1}{f_2(y)} \frac{\partial  \Phi}{\partial y}
           \,\bm{e}^2\wedge\bm{e}^{\hat{2}}\wedge\bm{e}^0 ~.} \label{eq:TorsionAII}
\end{equation}
This type of metrics includes the charged, rotating Kaluza--Klein black holes
in five-dimensional minimal supergravity discussed in the next section.

\subsection{Type B metric}
Let us consider the next case $M_1=M_2=0$.
This is an exceptional type appearing only when the torsion is present.
In this case, we find a solution
\begin{equation}
\eqalign{ 
 \bm{e}_\mu = \sqrt{\frac{X_\mu}{x_\mu^2-x_\nu^2}}
              \Big(\frac{\partial}{\partial x_\mu}-Z_\mu \frac{\partial}{\partial\psi}\Big) ~, \\
 \bm{e}_{\hat{\mu}} = \sqrt{\frac{Y_\mu}{x_\mu^2-x_\nu^2}}\frac{\partial}{\partial y_\mu} ~, \quad
 \bm{e}_0 = \frac{\partial}{\partial\psi} ~,
}
\end{equation}
where $X_\mu$, $Y_\mu$ and $Z_\mu$ are functions of two variables $x_\mu$ and $y_\mu$.
Rewriting $(x_1,x_2,y_1,y_2)=(x,y,\tau,\sigma)$,
we obtain the corresponding metric by
\begin{equation}
\fl \eqalign{
 \bm{g}\ =\ & \frac{x^2-y^2}{X_1(x,\tau)}dx^2+\frac{y^2-x^2}{X_2(y,\sigma)}dy^2
            + \frac{x^2-y^2}{Y_1(x,\tau)}d\tau^2 + \frac{y^2-x^2}{Y_2(y,\sigma)}d\sigma^2 \\
         & +\Big\{d\psi + Z_1(x,\tau)\,dx + Z_2(y,\sigma)\,dy\Big\}^2 ~.}
\end{equation}
The metric contains 6 unknown functions of two variables:
$X_1(x,\tau)$, $Y_1(x,\tau)$, $Z_1(x,\tau)$, $X_2(y,\sigma)$, $Y_2(y,\sigma)$ and $Z_2(y,\sigma)$.
The torsion is given by
\begin{equation}
\fl \eqalign{ \bm{T}\ =\ & \frac{2x}{x^2-y^2}\sqrt{\frac{X_2(y,\sigma)}{y^2-x^2}} 
            \,\bm{e}^1\wedge\bm{e}^{\hat{1}}\wedge\bm{e}^{\hat{2}}
           + \frac{2y}{y^2-x^2}\sqrt{\frac{X_1(x,\tau)}{x^2-y^2}}
            \,\bm{e}^2\wedge\bm{e}^{\hat{2}}\wedge\bm{e}^{\hat{1}} \\
         & +\sqrt{\frac{X_1(x,\tau)Y_1(x,\tau)}{(x^2-y^2)^2}}\frac{\partial Z_1(x,\tau)}{\partial \tau}
            \,\bm{e}^1\wedge\bm{e}^{\hat{1}}\wedge \bm{e}^0 \\
         & +\sqrt{\frac{X_2(y,\sigma)Y_2(y,\sigma)}{(y^2-x^2)^2}}\frac{\partial Z_2(y,\sigma)}{\partial \sigma}
            \,\bm{e}^2\wedge\bm{e}^{\hat{2}}\wedge \bm{e}^0 ~. }
\end{equation}

\subsection{Type C metric}
In the last case, by virtue of symmetry between $x_1$ and $x_2$, 
we can take $K_2=0$ and $M_2=0$ without loss of generality.
A solution we found is given by
\begin{equation}
\eqalign{
 \bm{e}_1 = \sqrt{\frac{X(x,\tau)}{x^2-y^2}}\frac{\partial}{\partial x} ~, \quad 
 \bm{e}_2 = \sqrt{\frac{Y(y)}{y^2-x^2}}\frac{\partial}{\partial y} ~, \quad
 \bm{e}_0 = \frac{\partial}{\partial\psi} ~, \\
 \bm{e}_{\hat{1}} = \frac{1}{\sqrt{x^2-y^2}\Psi_1(x)}\Big(
  \frac{\partial}{\partial \tau}+\Omega_1(x)\frac{\partial}{\partial \sigma}
  +\lambda\frac{\partial}{\partial \psi}\Big) ~, \\
 \bm{e}_{\hat{2}} = \frac{1}{\sqrt{x^2-y^2}\Psi_2(y)}\Big(
  \frac{\partial}{\partial \sigma}+\Omega_2(y)\frac{\partial}{\partial \psi}\Big) ~,
}
\end{equation}
which contains 5 functions of single variable $Y(y)$, $\Psi_1(x)$, $\Psi_2 (y)$, $\Omega_1(x)$, $\Omega_2(y)$,
and a function of two variables $X(x,\tau)$.
Since the $\lambda$ is a gauge parameter, we set $\lambda=0$.
The metric is
\begin{eqnarray}
\fl \eqalign{\bm{g}
  \ =\ &\frac{x^2-y^2}{X(x,\tau)}dx^2+\frac{y^2-x^2}{Y(y)}dy^2 
+\Big\{d\psi + \Omega_1(x)\Omega_2(y) d\tau-\Omega_2(y) d\sigma\Big\}^2 \\
 &  +(x^2-y^2)\Big\{\Psi_1(x)^2d\tau^2+\Psi_2 (y)^2(-\Omega_1(x) d\tau+d\sigma)^2\Big\} ~. }
\end{eqnarray}
The torsion is given by
\begin{eqnarray}
 \eqalign{\bm{T} =& \left(\frac{2x}{x^2-y^2}\sqrt{\frac{Y(y)}{y^2-x^2}}-\frac{\Psi_2(y)}{\Psi_1(x)}
            \frac{\partial\Omega_1(x)}{\partial x}\sqrt{\frac{X(x,\tau)}{x^2-y^2}}\right)
            \bm{e}^1\wedge\bm{e}^{\hat{1}}\wedge\bm{e}^{\hat{2}} \\
         & +\frac{2y}{y^2-x^2}\sqrt{\frac{X(x,\tau)}{x^2-y^2}}
           ~\bm{e}^2\wedge\bm{e}^{\hat{2}}\wedge\bm{e}^{\hat{1}} \\
         & +\frac{\Omega_2(y)}{\Psi_1(x)}\frac{\partial\Omega_1(x)}{\partial x}
            \sqrt{\frac{X(x,\tau)}{(x^2-y^2)^2}}
            ~\bm{e}^1\wedge\bm{e}^{\hat{1}}\wedge\bm{e}^0 \\
         & +\frac{1}{\Psi_2}\frac{\partial\Omega_2(y)}{\partial y}\sqrt{\frac{-Y(y)}{(x^2-y^2)^2}}
            ~\bm{e}^2\wedge\bm{e}^{\hat{2}}\wedge\bm{e}^0 \,. }
\end{eqnarray}

\section{Killing--Yano symmetry of Kaluza--Klein black holes in Einstein--\\ Maxwell--Chern--Simons theory}
In this section, we investigate Killing--Yano symmetry 
of Kaluza--Klein black holes in five-dimensional Einstein--Maxwell--Chern--Simons theory.
The action of the theory consists of a (Lorentzian) metric $g_{\mu\nu}$ and a Maxwell field $A_\mu$,
\begin{equation}
 S = \int *(R+\Lambda) -\frac{1}{2}* \bm{F}\wedge \bm{F}
          + \frac{\lambda_{cs}}{3\sqrt{3}}\bm{F}\wedge \bm{F}\wedge \bm{A} ~. \label{ActionMS}
\end{equation}
where $\bm{F}=d\bm{A}$ is a field strength of the Maxwell field,
$\Lambda$ is a cosmological constant and
$\lambda_{cs}$ is coupling constant of the Chern--Simons term.
This is said to be the pure Einstein--Maxwell theory when $\lambda_{cs}=0$
and the minimal supergravity when $\lambda_{cs}=1$.
The equations of motion are given by
\begin{eqnarray}
 R_{ab} + \frac{\Lambda}{3}g_{ab} = \frac{1}{2}\Big(F_{ac}F_b{}^c-\frac{1}{6}g_{ab}F_{cd}F^{cd}\Big) ~,
  \label{EinsteinEq} \\
 d* \bm{F} - \frac{\lambda_{cs}}{\sqrt{3}} \bm{F}\wedge \bm{F}= 0 ~.
  \label{MaxwellEq}
\end{eqnarray}

\subsection{Uplift of the Kerr--Newman solution}\label{sec:uplift}
Many exact solutions of  five-dimensional Einstein--Maxwell--Chern--Simons theory
have already been discovered in literature.
Of them, we focus especially on Kaluza--Klein type metrics,
\begin{equation}
 \bm{g} = e^{-2\phi/\sqrt{3}}\bm{g}^{(4)} + e^{\phi/\sqrt{3}}(d\psi + {\cal W}_i \,dx^i)^2 ~,
\end{equation}
with a Maxwell field written in the form
\begin{equation}
 \bm{A} = {\cal A}_i \,dx^i + \rho \,d\psi ~,
\end{equation}
where ${\cal W}_i$, ${\cal A}_i$, $\phi$ and $\rho$ are functions
of the coordinates $x^i$ in four dimensions.
Then we find that by setting
\begin{equation}
 \phi = \rho = 0 ~,~~~
 \star \bm{{\cal G}} = \frac{1}{\sqrt{3}} \bm{{\cal F}} ~, \label{costrun}
\end{equation}
where $\star$ represents the Hodge star with respect to $\bm{g}^{(4)}$,
$\bm{{\cal G}} = d\bm{{\cal W}}$ and $\bm{{\cal F}} = d\bm{{\cal A}}$
are field strengths in four dimensions,
the action (\ref{ActionMS}) consistently reduces
to the four-dimensional Einstein--Maxwell theory (e.g., see \cite{Elvang:2005}).
According to \cite{Matsuno:2012}, this identification leads to an uplift of the
Reissner--Nordstr\"om solution for an arbitrary value of the coupling $\lambda _{cs}$.
This fact motivates us to consider $S^1$ bundle over the Kerr--Newman spacetime.
That is, the ansatz is the following:
\begin{eqnarray}
\fl \eqalign{
 \bm{g} =& -\frac{\Delta}{\Sigma}(dt - a \sin^2\theta d\phi)^2
           +\frac{\Sigma}{\Delta}dr^2 + \Sigma d\theta^2
           +\frac{\sin^2\theta}{\Sigma}(a dt - (r^2+a^2)d\phi)^2  \\
    & +\Big(\frac{L}{2}d\psi - \frac{\alpha qr}{\Sigma}(dt - a \sin^2\theta d\phi) 
                          - \frac{\beta q\cos\theta}{\Sigma} (a dt - (r^2+a^2)d\phi)\Big)^2 ~, \label{met2} } \\
\fl \bm{A} = - \frac{\gamma qr}{\Sigma}(dt - a \sin^2\theta d\phi) 
                          - \frac{\delta q\cos\theta}{\Sigma} (a dt - (r^2+a^2)d\phi) ~, \label{maxwell2}
\end{eqnarray}
and
\begin{equation}
 \Delta = r^2 + a^2 + q^2 -2m r ~,~~~ \Sigma = r^2 + a^2\cos^2\theta ~.
\end{equation}
If the Einstein--Maxwell--Chern--Simons theory is imposed,
the equations of motion require
the parameters $\alpha$, $\beta$, $\gamma$ and $\delta$ to satisfy
some algebraic relations.
The Einstein's Eq.\ (\ref{EinsteinEq}) yields
\begin{eqnarray}
 \alpha^2+\beta^2+\gamma^2+\delta^2 = 4 ~, \\
 3(\alpha^2-\beta^2)+\gamma^2-\delta^2 = 0 ~,\\
 3\alpha \beta+\gamma\delta = 0 ~, \label{constraint3}
\end{eqnarray}
which can be solved by
\begin{equation}
 \alpha = \frac{\delta}{\sqrt{3}} ~,~~~ \beta = -\frac{\gamma}{\sqrt{3}} ~,~~~
 \alpha^2+\beta^2=1 ~. \label{rel2}
\end{equation}
These relations actually imply the conditions (\ref{costrun}).
Although the Maxwell's Eq.\  (\ref{MaxwellEq}) impose additional conditions
\begin{eqnarray}
&& \alpha\gamma-\beta\delta-\frac{2\lambda_{cs}}{\sqrt{3}} \gamma\delta = 0 ~, \\
&& \alpha\delta+\beta\gamma+\frac{\lambda_{cs}}{\sqrt{3}} (\gamma^2-\delta^2) = 0 ~,
\end{eqnarray}
the relations (\ref{rel2}) automatically guarantee them when $\lambda_{cs}=1$.
Otherwise, we have the trivial solution $\alpha=\beta=\delta=\gamma=0$.
If we were to set $a=0$ which corresponds to non-rotating (four-dimensional) 
spacetimes, Eq. (\ref{constraint3}) would have been absent and dynamical solutions 
could have been obtained for $\lambda_{cs} \neq 1$\cite{Matsuno:2012}. 
In the presence of rotation $a\neq 0$, only the minimal supergravity can 
accommodate the uplift within the present setup.

The metric describes charged, rotating Kaluza--Klein black holes when $\beta\neq 0$
and black strings when $\beta=0$.
The constant $L$ represents the size of the extra dimension.
The parameters $m$, $a$, $q$, $\alpha$, $\gamma$ and $\delta$
are related to five charges:
mass $M$, angular momentum $J^\psi$ and $J^\phi$ that are associated with the $\psi$ and $\phi$ directions, 
electric charge $Q$,  and magnetic flux $\Psi$.
Thus obtained solution is trivial because this is just an uplift of the four-dimensional 
Kerr--Newman solution. However, as we will see below,
this is an interesting example from the view point of generalised Killing--Yano symmetry.

\subsection{Hidden symmetry}
In this subsection, we show that the solutions obtained above fall into
the general class of metrics derived in the previous section. Then, by writing down the
Hamilton--Jacobi equation and Klein--Gordon equation on those spacetimes, we
explicitly demonstrate the connection between the separability of these equations
and the underlying Killing--Yano symmetry.

\subsection*{3.2.1. Killing--Yano symmetry}
The metric (\ref{met2}) admits three Killing vectors $\partial/\partial \tau$, 
$\partial/\partial \sigma$ and $\partial/\partial \psi$.
Besides them, we can find a rank-2 Killing--St\"ackel tensor
and a rank-2 generalised Killing--Yano tensor.
To see this, it is helpful to use the coordinates
\begin{equation}
 p = a\cos\theta ~,~~~ \tau = t - a \phi ~,~~~ \sigma = \frac{\phi}{a} ~, \label{coordtransf}
\end{equation}
which Carter introduced in \cite{Carter:1968}
to study separability of the Hamilton--Jacobi equation for geodesics in the Kerr spacetime.
The coordinate transformation (\ref{coordtransf})
brings the metric (\ref{met2}) to a simple algebraical form
\begin{eqnarray}
\eqalign{
 \bm{g} =&\frac{r^2+p^2}{{\cal Q}}dr^2 + \frac{r^2+p^2}{{\cal P}} dp^2  \\
       & -\frac{f_1^2\,{\cal Q}}{r^2+p^2}(d\tau +p^2 d\sigma)^2
           +\frac{f_2^2\,{\cal P}}{r^2+p^2}(d\tau - r^2 d\sigma)^2  \\
       & +\Big(d\psi + \frac{N_1}{r^2+p^2}(d\tau +p^2 d\sigma) 
                          + \frac{N_2}{r^2+p^2} (d\tau - r^2 d\sigma)\Big)^2 ~,} \label{met3}
\end{eqnarray}
where
\begin{eqnarray}
\eqalign{
& {\cal Q} = r^2 + a^2 + q^2 -2m r ~,~~~ {\cal P} = a^2- p^2 ~,  \\
& N_1 = -\alpha q r ~,~~~ N_2 = -\beta qp ~,~~~ f_1 = 1 ~,~~~ f_2 = 1 ~.} \label{choice1}
\end{eqnarray}
The inverse metric is given by
\begin{eqnarray}
\eqalign{
 \Big(\frac{\partial}{\partial s}\Big)^2
=& - \frac{1}{f_1^2(r^2+p^2){\cal Q}}
     \Big(r^2 \frac{\partial}{\partial \tau} 
          + \frac{\partial}{\partial \sigma}
          -N_1 \frac{\partial}{\partial \psi} \Big)^2  \\
 & + \frac{1}{f_2^2(r^2+p^2){\cal P}}
     \Big(p^2 \frac{\partial}{\partial \tau}
          - \frac{\partial}{\partial \sigma}
          -N_2 \frac{\partial}{\partial \psi} \Big)^2  \\
 & +\frac{{\cal Q}}{r^2+p^2}\Big(\frac{\partial}{\partial r}\Big)^2
   +\frac{{\cal P}}{r^2+p^2}\Big(\frac{\partial}{\partial p}\Big)^2
   +\Big(\frac{\partial}{\partial \psi}\Big)^2 ~,} \label{inverse1}
\end{eqnarray}

Looking at (\ref{met3}), we notice that the metric is a Lorentzian counterpart
of the type A$_\infty$ metric (\ref{metAII}) obtained in Sec.\ 2.4.
This means that (\ref{met3}) admits a rank-2 GKY tensor.
We may consider the metric with the functions of (\ref{choice1}) replaced
by arbitrary one variable functions ${\cal Q}(r)$, ${\cal P}(p)$, $N_1(r)$, $N_2(p)$,
$f_1(r)$ and $f_2(p)$, which we call off-shell metric.
By considering such an off-shell metric,
we can deal with the metric (\ref{met2}) in a more algebraically general framework.
For the off-shell metric (\ref{met3}), the canonical orthonormal frame is introduced as 
\begin{eqnarray}
\eqalign{
& \bm{e}^1 = \sqrt{\frac{r^2+p^2}{{\cal Q}}}dr ~,~~~ 
  \bm{e}^{\hat{1}} = f_1\sqrt{\frac{{\cal Q}}{r^2+p^2}}(d\tau +p^2 d\sigma) ~,  \\
& \bm{e}^2 = \sqrt{\frac{r^2+p^2}{{\cal P}}}dp ~,~~~
  \bm{e}^{\hat{2}} = f_2\sqrt{\frac{{\cal P}}{r^2+p^2}}(d\tau - r^2 d\sigma) ~,  \\
& \bm{e}^0 = d\psi + \frac{N_1}{r^2+p^2}(d\tau +p^2 d\sigma) 
                          + \frac{N_2}{r^2+p^2} (d\tau - r^2 d\sigma) ~.}
\end{eqnarray}
With respect to this canonical frame, we can easily write down
the metric and the rank-2 GKY tensor as
\begin{eqnarray}
 \bm{g} &=& \bm{e}^1\bm{e}^1 -\bm{e}^{\hat{1}}\bm{e}^{\hat{1}}
             + \bm{e}^2\bm{e}^2+\bm{e}^{\hat{2}}\bm{e}^{\hat{2}}+\bm{e}^0\bm{e}^0 ~, \label{met9} \\
 \bm{f} &=& p \,\bm{e}^1\wedge\bm{e}^{\hat{1}} + r \,\bm{e}^2\wedge \bm{e}^{\hat{2}} ~. \label{KY9}
\end{eqnarray}
From (\ref{KYtoKT}), we obtain a rank-2 Killing--St\"ackel tensor
\begin{equation}
 \bm{K} = p^2 (\bm{e}^1\bm{e}^1-\bm{e}^{\hat{1}}\bm{e}^{\hat{1}})
          -r^2(\bm{e}^2\bm{e}^2+\bm{e}^{\hat{2}}\bm{e}^{\hat{2}}) ~, \label{KT9}
\end{equation}
which is given in terms of the coordinate basis by
\begin{eqnarray}
\eqalign{
 K^{ab}\frac{\partial}{\partial x^a}\frac{\partial}{\partial x^b}
=& - \frac{p^2}{f_1^2(r^2+p^2){\cal Q}}
     \Big(r^2 \frac{\partial}{\partial \tau} 
          + \frac{\partial}{\partial \sigma}
          -N_1 \frac{\partial}{\partial \psi} \Big)^2  \\
 & + \frac{r^2}{f_2^2(r^2+p^2){\cal P}}
     \Big(p^2 \frac{\partial}{\partial \tau}
          - \frac{\partial}{\partial \sigma}
          -N_2 \frac{\partial}{\partial \psi} \Big)^2  \\
 & +\frac{p^2{\cal Q}}{r^2+p^2}\Big(\frac{\partial}{\partial r}\Big)^2
   -\frac{r^2{\cal P}}{r^2+p^2}\Big(\frac{\partial}{\partial p}\Big)^2 ~.} \label{KT1}
\end{eqnarray}

\subsection*{3.2.2. Separation of variables in the Hamilton--Jacobi equation}
The existence of rank-2 Killing--St\"ackel tensors is in general related 
to separation of variables in Hamilton--Jacobi equations for geodesics
\begin{equation}
 g^{ab} \partial_a S \partial_b S = -m^2 ~. \label{GeodesicEq}
\end{equation}
For the off-shell metric (\ref{met3}), since the inverse metric is given by (\ref{inverse1}),
the Hamilton--Jacobi equation for geodesics (\ref{GeodesicEq}) can be solved by separation of variables
with a function
\begin{equation}
 S = R(r) + \Theta(p) + \pi_\tau \tau + \pi_\sigma\sigma + \pi_\psi\psi ~, \label{sov1}
\end{equation}
where $\pi_\tau$, $\pi_\sigma$ and $\pi_\psi$ are arbitrary constants, and 
the functions $R(r)$ and $\Theta(p)$ satisfy the ordinary differential equations
\begin{equation}
 \Big(\frac{dR}{dr}\Big)^2 - \Big(\frac{W_r}{f_1{\cal Q}}\Big)^2 - \frac{V_r}{{\cal Q}} =0 ~,~~~
 \Big(\frac{d\Theta}{dp}\Big)^2 + \Big(\frac{W_p}{f_2{\cal P}}\Big)^2 - \frac{V_p}{{\cal P}} = 0 \label{sov2}
\end{equation}
with the potentials including a separation constant $\kappa$,
\begin{eqnarray}
\eqalign{
 W_r = r^2\pi_\tau+ \pi_\sigma - N_1 \pi_\psi ~,~~~
 V_r = -(\pi_\psi^2+m^2)r^2+\kappa ~,  \\
 W_p = p^2\pi_\tau- \pi_\sigma - N_2 \pi_\psi ~,~~~
 V_p = -(\pi_\psi^2+m^2)p^2-\kappa ~.} \label{potentials}
\end{eqnarray}
The constant $\kappa$ is given, by eliminating $-m^2$, as
\begin{equation}
 \kappa = \frac{p^2}{r^2+p^2}\Big({\cal Q}\Pi_r^2-\frac{W_r^2}{f_1^2{\cal Q}}\Big)
          -\frac{r^2}{r^2+p^2}\Big({\cal P}\Pi_p^2-\frac{W_p^2}{f_2^2{\cal P}}\Big) ~, \label{sov4}
\end{equation}
where $\Pi_a = \partial_a S$ is the canonical momentum given by
\begin{equation}
 \Pi_r = \pm\sqrt{\Big(\frac{W_r}{f_1{\cal Q}} \Big)^2 + \frac{V_r}{{\cal Q}}} ~,~~~
 \Pi_p = \pm\sqrt{-\Big(\frac{W_p}{f_2{\cal P}} \Big)^2 + \frac{V_p}{{\cal P}}}
\end{equation}
and $\Pi_\tau =\pi_\tau$, $\Pi_\sigma=\pi_\sigma$ and $\Pi_\psi=\pi_\psi$.

We have now four constants of motion.
Three of them $\Pi_\tau$, $\Pi_\sigma$ and $\Pi_\psi$
are associated with the Killing vectors $\partial/\partial \tau$, 
$\partial/\partial \sigma$ and $\partial/\partial \psi$.
The Killing--St\"ackel tensor $K_{ab}$ is responsible for the other one $\kappa$.
In fact, from the equation (\ref{KT1}), one can easily confirm that
\begin{equation}
 \kappa=K^{ab} \Pi_a \Pi_b ~ \label{KTtokappa} ,
\end{equation}
which is precisely the general relation (\ref{constantFromKS}). Its constancy is in general
guaranteed by the commutativity with the Hamiltonian $H = g^{ab}\Pi _a \Pi _b $ under
the Poisson bracket, namely $\{ H ,\kappa \} =0$.
Thus we have found that the charged, rotating Kaluza--Klein black hole spacetime (\ref{met2})
does posses the separable structure for the Hamilton--Jacobi equation and its separation
constant is indeed related to the underlying Killing--Yano symmetry that generates the rank-2
Killing--St\"ackel tensor necessary for separation of variables.

\subsection*{3.2.3 Separation of variables in the Klein-Gordon equation}
The massive Klein--Gordon equation is given by
\begin{equation}
 (\Box + m^2) \Psi = 0 ~. \label{KGeq}
\end{equation}
With the help of $\Box\Psi = (1/\sqrt{-g})\partial_a(\sqrt{-g}g^{ab}\partial_b\Psi)$, we have
\begin{eqnarray}
\fl \eqalign{
& \Bigg[\frac{1}{f_1}\frac{\partial}{\partial r}\Big(f_1{\cal Q}\frac{\partial }{\partial r}\Big)
 + \frac{1}{f_2}\frac{\partial}{\partial p}\Big(f_2{\cal P}\frac{\partial }{\partial p}\Big)
  - \frac{1}{f_1^2{\cal Q}}
     \Big(r^2 \frac{\partial}{\partial \tau} 
          + \frac{\partial}{\partial \sigma}
          -N_1 \frac{\partial}{\partial \psi} \Big)^2  \\
& + \frac{1}{f_2^2{\cal P}}
     \Big(p^2 \frac{\partial}{\partial \tau}
          - \frac{\partial}{\partial \sigma}
          -N_2 \frac{\partial}{\partial \psi} \Big)^2
    + (r^2+p^2)\Big(\frac{\partial^2}{\partial \psi^2} + m^2\Big)\Bigg]\Psi = 0 ~.}
\end{eqnarray}
This equation is separable with a function
\begin{equation}
 \Psi = R(r)\Theta(p) e^{\pi_\tau \tau + \pi_\sigma \sigma + \pi_\psi \psi} ~,
\end{equation}
where $\pi_\tau$, $\pi_\sigma$ and $\pi_\psi$ are constants again,
and the functions $R(r)$ and $\Theta(p)$ satisfy the ordinary differential equations
\begin{equation}
\fl  \frac{1}{f_1}\frac{d}{dr}\Big(f_1{\cal Q}\frac{dR}{dr}\Big)
 -\frac{W_r^2}{f_1^2{\cal Q}} - V_r = 0 ~,~~~
 \frac{1}{f_2}\frac{d}{dp}\Big(f_2{\cal P}\frac{d\Theta}{dp}\Big)
 +\frac{W_p^2}{f_2^2{\cal P}} - V_p = 0
\end{equation}
with the potentials given by (\ref{potentials}) including the separation constant $\kappa$.
By eliminating $m^2$, the constant $\kappa$ is this time given by
\begin{eqnarray}
\eqalign{
 \kappa =& \frac{p^2}{r^2+p^2}\Bigg[\frac{1}{f_1}\frac{d}{dr}\Big(f_1{\cal Q}\frac{dR}{dr}\Big)
 -\frac{W_r^2}{f_1^2{\cal Q}}\Bigg]  \\
         &+\frac{r^2}{r^2+p^2}\Bigg[\frac{1}{f_2}\frac{d}{dp}\Big(f_2{\cal P}\frac{d\Theta}{dp}\Big)
 +\frac{W_p^2}{f_2^2{\cal P}}\Bigg] ~.}
\end{eqnarray}
Again, it is straightforward to check that this constant coincides with the one following
from the general property of the symmetry operator (\ref{symmetryOperator})
together with the Killing--St\"ackel tensor (\ref{KT9}), that is, we have
\begin{equation}
 \hat{K}\Psi \equiv \nabla_a K^{ab} \nabla_b \Psi = \kappa \Psi ~.
\end{equation}
This is the relationship between Killing--St\"ackel tensor
and separability of the Klein--Gordon equation.

\subsection{More general solution}
Our purpose here is to construct the most general solution of 
the Einstein-Maxwell-Chern-Simons theory, namely the equations (\ref{EinsteinEq})
and (\ref{MaxwellEq}), for the type A$_\infty$ off-shell metric (\ref{met3}).
We adopt the ansatz that identifies torsion with the flux by
\begin{equation}
 \bm{T} = \frac{1}{\sqrt{3}}*\bm{F} ~. \label{TorsionMinimal}
\end{equation}

Under the assumption (\ref{TorsionMinimal}), the Maxwell-Chern-Simons equation 
(\ref{MaxwellEq}) and the Bianchi identity, $d\bm{F}=0$, are written\footnote{
It was shown in \cite{Ahmedov:2009} that a five-dimensional spacetime
admitting a rank-2 closed CKY tensor which is a generalised closed CKY tensor at the same time
with the torsion satisfying (\ref{Eqmaxwell}) is uniquely given
by the Chong-Cveti\v{c}-L\"u-Pope solution \cite{Chong:2005} in five-dimensional minimal supergravity.
} as
\begin{equation}
 d \bm{T} - \lambda_{cs} (*\bm{T}) \wedge (*\bm{T}) = 0 ~,~~~
 d * \bm{T} = 0 ~. \label{Eqmaxwell}
\end{equation}
Substituting the expression for the torsion (\ref{eq:TorsionAII}),
the first equation requires both $f_1 = f_2$ and $\lambda_{cs} = 1$.
Since $f_1$ and $f_2$ are one variable functions of only $r$ and $p$ respectively,
we find that $f_1=f_2$ must be constant
and then it can be absorbed in $N_\mu$ via rescaling of $\tau$ and $\sigma$.
Thus, we may set $f_1=f_2=1$.
The restriction to $\lambda_{cs}=1$ corresponds to the minimal supergravity.
The second equation then solves as
\begin{equation}
 N_1 = \tilde{a} r^2 + b_1 r ~,~~~ N_2 = \tilde{a} p^2 +b_2 p ~,
\end{equation}
where $\tilde{a}$, $b_1$ and $b_2$ are constants. 
Note that $\tilde{a}$ is a gauge parameter 
which can be eliminated by gauge transformation of $\psi$.
We set it to be zero.
These conditions render many
components of the Einstein equations trivial. We need $(0,0)$ component to
derive $\Lambda = 0$. Then $(3,3)$ and $(4,4)$ components determine
$\mathcal{Q}$ and $\mathcal{P}$ as
\begin{equation}
\mathcal{Q} = \tilde{c} r^2 + m_1 r + q_1~, \qquad \mathcal{P} = -\tilde{c} p^2 + m_2 p + q_2 ~, 
\end{equation}
where $\tilde{c} , m_1 , m_2 , q_1$ and $q_2$ are constants which satisfy
\begin{equation}
q_1 -q_2 =  b_1^2 + b_2^2   .
\end{equation}
Again the gauge freedom enables us to rescale $\tilde{c}$ to be $1$ so that we have 
obtained a five-parameter family of solutions.

In order to compare it with the Kaluza--Klein black hole solution in Sec. \ref{sec:uplift},
we make the coordinate transformation (\ref{coordtransf}) and set $\mu = m_2 /a$ 
and $q_0 = q_2 /a^2$. 
The obtained solution is written as
\begin{equation}
\fl \eqalign{
\bm{g} = &-\frac{\Delta }{\Sigma } \left( dt -a\sin ^2 \theta d\phi \right) ^2 + \frac{\Sigma }{\Delta } dr^2 
+ \frac{\Sigma \sin ^2 \theta }{\sin ^2 \theta + \mu \cos \theta + q_0 -1} d\theta ^2 \\
& + \frac{\sin ^2 \theta + \mu \cos \theta + q_0 -1  }{\Sigma } \left( a dt - \left( r^2 + a^2 \right) d\phi \right) ^2 \\
& + \left( d\psi + \frac{b_1 r}{\Sigma } \left( dt - a\sin ^2 \theta d\phi \right) + \frac{b_2 \cos \theta }{\Sigma } \left( a dt - \left( r^2 +a^2 \right) d\phi \right) \right) ^2 ~, \\
\bm{A} = & - \frac{\sqrt{3}b_1 r}{\Sigma } \left( dt - a\sin ^2 \theta d\phi \right) + \frac{\sqrt{3}b_2 \cos \theta }{\Sigma } \left( adt - \left( r^2 + a^2 \right) d\phi \right) ~,}
\end{equation}
where
\begin{equation}
\Delta = r^2 + m_1 r + a^2 q_0 + b_1^2 + b_2^2 ~, \qquad \Sigma = r^2 + a^2 \cos ^2 \theta ~.
\end{equation}
It can be seen that the metric is the uplift of the Kerr--Newman--NUT solution \cite{Carter:1968}.
The result of Sec. \ref{sec:uplift} is included as a special case $\mu = 0$ and $q_0 =1$.

\section{Construction of Kaluza--Klein black holes in heterotic supergravity}
In this section we consider abelian heterotic supergravity in five dimensions, 
which is the low-energy effective theory of heterotic string theory.
The string-frame action consists of a (Lorentzian) metric $g_{\mu\nu}$, scalar field $\varphi$,
$U(1)$ gauge potential $A_\mu$ and 2-form potential $B_{\mu\nu}$,
\begin{equation}
 S = \int e^\varphi \Big( R + *d\varphi\wedge d\varphi 
     - *\bm{F}\wedge \bm{F} -\frac{1}{2}*\bm{H}\wedge \bm{H} \Big) ~,
\end{equation}
where $\bm{F} = d\bm{A}$ and $\bm{H}=d\bm{B} - \bm{A}\wedge d\bm{A}$.
The equations of motion are
\begin{eqnarray}
&& R_{ab}-\nabla_a\nabla_b \varphi -F_a{}^cF_{bc}-\frac{1}{4}H_a{}^{cd}H_{bcd} = 0 ~, 
  \label{HeteroEinsteinEq} \\
&& d(e^{\varphi}*\bm{F})-e^{\varphi} *\bm{H} \wedge \bm{F} = 0 ~, 
  \label{HeteroMaxwellEq} \\
&& d(e^{\varphi}*\bm{H})= 0 ~,
  \label{HeteroThreeFormEq}\\
&& R - (\nabla\varphi)^2 - 2\nabla^2\varphi - \frac{1}{2}F^2 -\frac{1}{12}H^2 = 0 ~.
  \label{HeteroScalarEq}
\end{eqnarray}
We start from reviewing a known black string solution of this theory, revealing
its Killing--Yano symmetry. Then using the general form of metric obtained in Sec.\ 2, 
we construct a class of charged, rotating Kaluza--Klein black holes.

\subsection{Hidden symmetry of the Mahapatra's solution}
Using the technique of S.\ F.\ Hassan and A.\ Sen \cite{Hassan:1992},
a charged, rotating black string solution 
with four parameters $(m,a,\delta_1,\delta_2)$ was constructed by S.\ Mahapatra \cite{Mahapatra:1994}.
The solution is given by
\begin{eqnarray}
 \bm{g} &=& -\frac{\Delta}{\Sigma}\Big(dt -a \sin^2\theta d\phi - \bm{W}\Big)^2
         + \frac{a^2\sin^2\theta}{\Sigma}\Big(dt -\frac{r^2+a^2}{a} d\phi - \bm{W}\Big)^2 \nonumber \\
       && +\frac{\Sigma}{\Delta}dr^2 +\Sigma d\theta^2 + \Big(d\psi - \frac{\beta}{1-\alpha} \bm{W}\Big)^2 ~,
 \label{MahapatraMetric}\\
 \bm{A} &=& -\frac{\gamma}{1-\alpha} \bm{W} ~, 
 \label{MahapatraOneForm}\\
 \bm{B} &=& \Big(-dt + \frac{\beta}{1-\alpha}d\psi\Big)\wedge \bm{W} ~, 
 \label{MahapatraTwoForm}\\
 e^\varphi &=& 1 + \frac{-mr(1-\alpha)}{\Sigma} ~,
 \label{MahapatraScalar}
\end{eqnarray}
where
\begin{eqnarray}
&& \Delta = r^2 + a^2 -2mr ~,~~~ \Sigma = r^2+a^2\cos^2\theta ~, \\
&& \bm{W} = \frac{-mr(1-\alpha)}{\Sigma-mr(1-\alpha)}(dt -a \sin^2\theta d\phi) ~. \label{MahapatraW}
\end{eqnarray}
The parameters $\alpha$, $\beta$ and $\gamma$ are required to satisfy 
the relation $\alpha^2=1+\beta^2+\gamma^2$,
which can be written by two parameters $\delta_1$ and $\delta_2$ as
\begin{equation}
 \alpha = \cosh\delta_1 \cosh\delta_2 ~,~~~ \beta = \cosh\delta_1\sinh\delta_2 ~,~~~ 
 \gamma = \sinh\delta_1 ~.
\end{equation}

The field strengths $\bm{F}$ and $\bm{H}$ are easily computed as
\begin{equation}
 \bm{F} = -\frac{\gamma}{1-\alpha} d\bm{W} ~,~~~
 \bm{H} = \bm{\eta} \wedge d \bm{W} ~,
\end{equation}
where
\begin{eqnarray}
\bm{\eta} &=
& dt - \frac{\beta}{1-\alpha} d\psi - \frac{\gamma^2}{(1-\alpha)^2} \bm{W} \nonumber \\
&=& \frac{\Delta}{\Sigma}(dt -a \sin^2\theta d\phi - \bm{W})
  -\frac{a^2\sin^2\theta}{\Sigma}(dt -\frac{r^2+a^2}{a} d\phi - \bm{W}) \nonumber \\
&& -\frac{\beta}{1-\alpha}(d\psi-\frac{\beta}{1-\alpha}\bm{W}) ~.
\end{eqnarray}

\subsection*{4.1.1. Killing--Yano symmetry}
Let us unveil the Killing--Yano symmetry of the Mahapatra's spacetime.
Performing the coordinate transformation (\ref{coordtransf}) again,
the metric (\ref{MahapatraMetric}) can be written 
in a simple algebraical form
\begin{eqnarray}
 \bm{g} &=& -\frac{f_1^2{\cal Q}}{r^2+p^2}\Big(d\tau +p^2 d\sigma - \bm{W}\Big)^2
           + \frac{f_2^2{\cal P}}{r^2+p^2}\Big(d\tau -r^2 d\sigma - \bm{W}\Big)^2 \nonumber \\
         && +\frac{r^2+p^2}{{\cal Q}}dr^2 +\frac{r^2+p^2}{{\cal P}} dp^2
           +\Big(d\psi + \lambda \bm{W}\Big)^2 ~, 
 \label{MahapatraMetric2}
\end{eqnarray}
where
\begin{eqnarray}
\eqalign{
& \bm{W} = \frac{1}{\Phi}\Big(\frac{N_1}{\Sigma}(d\tau +p^2 d\sigma)
           + \frac{N_2}{\Sigma}(d\tau -r^2 d\sigma)\Big) ~,  \\
& \Phi = 1 + \frac{N_1}{\Sigma} + \frac{N_2}{\Sigma} ~,~~~
  \Sigma = r^2 + p^2 ~,  \\
& {\cal Q} = r^2 -2mr +a^2 ~,~~~ {\cal P} = a^2-p^2 ~,~~~
  N_1=-mr(1-\alpha) ~,  \\
& N_2=0 ~,~~~ f_1=1 ~,~~~ f_2 =1 ~,~~~  \lambda = - \frac{\beta}{1-\alpha} ~.} \label{defMetric2}
\end{eqnarray}
We observe that this metric falls into the family A$_\lambda$ (\ref{metAI}) 
in the general classification of Sec. 2.
Similarly to the previous section, we consider hidden symmetry
for the off-shell metric (\ref{MahapatraMetric2})
with ${\cal Q}$, ${\cal P}$, $N_1$, $N_2$, $f_1$, $f_2$ and $\lambda$
replaced by unknown functions ${\cal Q}(r)$, ${\cal P}(p)$, $N_1(r)$, $N_2(p)$,
$f_1 (r)$ and $f_2 (p)$ and an arbitrary constant $\lambda$.
For the off-shell metric (\ref{MahapatraMetric2}), the canonical frame is introduced as 
\begin{eqnarray}
\eqalign{
& \bm{e}^1 = \sqrt{\frac{r^2+p^2}{{\cal Q}}}dr ~,~~~ 
  \bm{e}^{\hat{1}} = f_1\sqrt{\frac{{\cal Q}}{r^2+p^2}}(d\tau +p^2 d\sigma- \bm{W}) ~,  \\
& \bm{e}^2 = \sqrt{\frac{r^2+p^2}{{\cal P}}}dp ~,~~~
  \bm{e}^{\hat{2}} = f_2\sqrt{\frac{{\cal P}}{r^2+p^2}}(d\tau - r^2 d\sigma -\bm{W}) ~,  \\
& \bm{e}^0 = d\psi + \lambda \bm{W} ~.}
\end{eqnarray}
With respect to this orthonormal frame, the metric
and the rank-2 generalised Killing--Yano tensor are written as (\ref{met9}) and (\ref{KY9}).
Furthermore, we also obtain a Killing--St\"ackel tensor of the form (\ref{KT9}).
Since the inverse metric is given by
\begin{eqnarray}
 \Big(\frac{\partial}{\partial s}\Big)^2
&=& - \frac{1}{f_1^2(r^2+p^2){\cal Q}}
     \Big((r^2+N_1)\frac{\partial}{\partial \tau} 
          + \frac{\partial}{\partial \sigma}
          - \lambda N_1\frac{\partial}{\partial \psi} \Big)^2 \nonumber \\
& & + \frac{1}{f_2^2(r^2+p^2){\cal P}}
     \Big((p^2+N_2)\frac{\partial}{\partial \tau}
          - \frac{\partial}{\partial \sigma}
           - \lambda N_2 \frac{\partial}{\partial \psi} \Big)^2 \nonumber \\
 && +\frac{{\cal Q}}{r^2+p^2}\Big(\frac{\partial}{\partial r}\Big)^2
   +\frac{{\cal P}}{r^2+p^2}\Big(\frac{\partial}{\partial p}\Big)^2
   +\Big(\frac{\partial}{\partial \psi}\Big)^2 ~,
\end{eqnarray}
the contravariant Killing--St\"ackel tensor can be written as
\begin{eqnarray}
\fl \eqalign{ K^{ab}\frac{\partial}{\partial x^a}\frac{\partial}{\partial x^b}
=& - \frac{p^2}{f_1^2(r^2+p^2){\cal Q}}
     \Big((r^2+N_1)\frac{\partial}{\partial \tau} 
          + \frac{\partial}{\partial \sigma}
          - \lambda N_1\frac{\partial}{\partial \psi} \Big)^2 \\
 & + \frac{r^2}{f_2^2(r^2+p^2){\cal P}}
     \Big((p^2+N_2)\frac{\partial}{\partial \tau}
          - \frac{\partial}{\partial \sigma}
           - \lambda N_2 \frac{\partial}{\partial \psi} \Big)^2 \\
 & +\frac{p^2{\cal Q}}{r^2+p^2}\Big(\frac{\partial}{\partial r}\Big)^2
   +\frac{r^2{\cal P}}{r^2+p^2}\Big(\frac{\partial}{\partial p}\Big)^2 ~. } \label{KT2}
\end{eqnarray}

\subsection*{4.1.2. Separation of variables in the Hamilton--Jacobi equation}
For the off-shell metric (\ref{MahapatraMetric2}), 
the geodesic equation (\ref{GeodesicEq}) can be solved 
by separation of variables with a function (\ref{sov1})
and obtain the ordinary differential equations (\ref{sov2})
with the potentials
\begin{eqnarray}
 W_r = (r^2+N_1)\pi_\tau+ \pi_\sigma - \lambda N_1 \pi_\psi ~,~~~
 V_r = -(\pi_\psi^2+m^2)r^2+\kappa ~, \nonumber \\
 W_p = (p^2+N_2)\pi_\tau- \pi_\sigma - \lambda N_2 \pi_\psi ~,~~~
 V_p = -(\pi_\psi^2+m^2)p^2-\kappa ~, \label{sov5}
\end{eqnarray}
where the separation constant $\kappa$ is given by (\ref{sov4})
with the above potentials.
Indeed, since this constant is related to the Killing--St\"ackel tensor (\ref{KT2})
with the general formula (\ref{KTtokappa}), we find that separation of variables 
for the geodesic equation in the Mahapatra's spacetime is underwritten by the existence of 
the generalised Killing--Yano symmetry.

\subsection*{4.1.3. Separation of variables in the Klein--Gordon equation}
In contrast to the case of minimal supergravity black holes,
the Klein--Gordon equation (\ref{KGeq}) for the off-shell metric (\ref{MahapatraMetric2})
does not separate.
On the other hand, the deformed Klein--Gordon equation
\begin{equation}
 (\Box + m^2) \Psi - (\nabla_a \varphi)(\nabla^a \Psi) = 0 \label{deformeKGeq2}
\end{equation}
does with $\varphi=\ln \Phi$.
Eq.\ (\ref{deformeKGeq2}) is equivalent to Klein--Gordon equation
in Einstein frame,
\begin{equation}
 (\Box_E + m^2 )\Psi = 0 ~,
\end{equation}
where $\Box_E$ is the d'Alembertian
with respect to the Einstein-frame metric $\bm{g}_E = \varphi^{2/3} \bm{g}$.
Namely, separation of variables for the Klein--Gordon equation
naturally occurs in the Einstein frame.
The similar situation was seen in the Kerr--Sen black holes \cite{Houri:2010}.

\subsection{Kaluza--Klein black hole solutions}
We attempt to find the general solutions of the equations of motion (\ref{HeteroEinsteinEq})--(\ref{HeteroScalarEq}) which take the form of type A$_\lambda$ metric (\ref{MahapatraMetric2}), under
the ansatz for the matter fields
\begin{equation}
 \varphi = \ln \Phi ~,~~~
 \bm{F} = c_F \,d\bm{W} ~,~~~
 \bm{H} = c_H\bm{T} ~,
\end{equation}
where $\bm{T}$ is the type A$_\lambda$ torsion (\ref{eq:torsionAI}).
For a non-trivial solution, equation (\ref{HeteroMaxwellEq}) requires
\begin{equation}
c_H = 1 ~.
\end{equation}
Under this condition, the rest of (\ref{HeteroMaxwellEq}) become dependent on the
dynamical equations of $\bm{H}$ (\ref{HeteroThreeFormEq}). Combined with the 
diagonal part of the Einstein equations (\ref{HeteroEinsteinEq}), one can derive
\begin{equation}
f_1 = f_2 ~, \qquad {\rm or} \qquad f_1 = \tilde{f} x^2 , \quad f_2 = \tilde{f} y^2 ~, \label{solutionf}
\end{equation}
where $\tilde{f} $ is constant. 
Since the latter is not consistent with the off-diagonal terms of the Einstein equations,
we focus on the former and set $f_1 = f_2 = 1$ by redefining the coordinates as before. 
The remaining component of
(\ref{HeteroThreeFormEq}) and the $(3,4)$-component of (\ref{HeteroEinsteinEq}) lead to
\begin{equation}
N_1 = \tilde{a} r^2 + b_1 r  , \quad N_2 = \tilde{a} p^2 + b_2 p  .\label{solutionN}
\end{equation}
Finally, the consistency condition $d\bm{H} = - \bm{F} \wedge \bm{F}$ determines 
$\mathcal{P}$ and $\mathcal{Q}$ as
\begin{equation}
\mathcal{P}(p) - \mathcal{Q}(r) = - \left( c_F^2 + \lambda ^2 \right) (r^2 + p^2 ) + c_{\Phi } (r^2 + p^2 )\Phi . \label{solutionP}
\end{equation}
(\ref{solutionf}), (\ref{solutionN}) and (\ref{solutionP}) are sufficient to guarantee that all the remaining equations are satisfied. 
By means of the physically irrelevant rescaling of $\Phi $, one can choose $\tilde{a} =0$ (note our
definition of $\Phi $ (\ref{defMetric2}) contains the normalised constant term) and derive
\begin{equation}
\eqalign{
\mathcal{Q} &= \left( c_F^2 + \lambda ^2 - c_{\Phi } \right) r^2 - b_1 c_{\Phi } r + c_0, \\
 \mathcal{P} &= - \left( c_F^2 + \lambda ^2 -c_{\Phi } \right) p^2 + b_2 c_{\Phi } p  +c_0 . } 
\end{equation}
The overall factor of $\mathcal{P}$ and $\mathcal{Q}$ can be gauged away, leaving
a family of solutions with five parameters $(c_0 , c_F , c_{\Phi } , b_1 , b_2 )$,
describing charged, rotating black holes and black strings.

Finally, by performing the coordinate transformation (\ref{coordtransf})
and setting $\mu=b_2/a$ and $q_0=c_0/a^2$,
the solution is written in the original coordinates as
\begin{eqnarray}
 \bm{g} &=& -\frac{\Delta}{\Sigma}\Big(dt -a \sin^2\theta d\phi - \bm{W}\Big)^2
         + \frac{a^2 Y}{\Sigma}\Big(dt -\frac{r^2+a^2}{a} d\phi - \bm{W}\Big)^2 \nonumber \\
       && +\frac{\Sigma}{\Delta}dr^2 +\frac{\Sigma \sin^2\theta}{Y} d\theta^2 
         + (d\psi + \lambda \bm{W})^2 ~,
 \label{MahapatraMetricNew}\\
 \bm{A} &=& c_F \bm{W} ~, 
 \label{MahapatraOneFormNew}\\
 \bm{B} &=& (-dt - \lambda d\psi)\wedge \bm{W} ~, 
 \label{MahapatraTwoForm}\\
 e^\varphi &=& 1 + \frac{b_1 r + \mu a^2 \cos\theta}{\Sigma} ~,
 \label{MahapatraScalarNew}
\end{eqnarray}
where
\begin{eqnarray}
 \Delta &=& (c_F^2+\lambda^2-c_\Phi)r^2 - b_1c_\Phi r + a^2 q_0 ~,~~~ 
 \Sigma = r^2+a^2\cos^2\theta ~, \nonumber\\
 Y &=& (c_F^2+\lambda^2-c_\Phi)\sin^2\theta+\mu c_\Phi \cos\theta + q_0-c_F^2-\lambda^2+c_\Phi ~, \nonumber\\
 \bm{W} &=& \frac{b_1 r}{\Sigma +b_1 r + \mu a^2 \cos\theta}(dt -a \sin^2\theta d\phi) \nonumber\\
  &&+ \frac{\mu \cos\theta}{\Sigma +b_1 r + \mu a^2 \cos\theta}(a dt -(r^2+a^2) d\phi) ~.
\end{eqnarray}
It is easy to check that the Mahapatra's solution (\ref{MahapatraMetric})--(\ref{MahapatraW}) is
recovered when we take $\lambda=-\beta/(1-\alpha)$, $c_F=-\gamma/(1-\alpha)$, $c_\Phi=-2/(1-\alpha)$,
$b_1=-m(1-\alpha)$, $\mu=0$ and $q_0 = c_F^2+\lambda^2-c_\Phi=1$.
It should be commented that the present metric can be regarded as 
an uplift of the Kerr--Sen--NUT solution obtained by \cite{Houri:2010}.

\section{Summary and Discussions}
Firstly, we have classified five-dimensional metrics admitting 
a rank-2 generalised Killing--Yano tensor
under the assumption that its eigenvector associated
with the zero eigenvalue is a Killing vector field.
The metrics have been classified into three types A, B and C in general,
and local expressions of the corresponding metrics have been obtained explicitly.
One of the open problems is to classify them without assuming the Killing vector,
in the presence of torsion. In this case, the large number of unknown variables arising
from the components of $\bm{\xi}$ makes it difficult to solve the integrability conditions.
However, even if it is not possible to obtain a general classification, it would be of a
great interest as an attempt to find spacetimes of less homogeneity.

We also have demonstrated separability structures of the Hamilton--Jacobi 
and Klein--Gordon equations for some known charged, rotating Kaluza--Klein black hole 
and black string solutions in the five-dimensional minimal supergravity
as well as heterotic supergravity. We have found that those spacetimes fall into the
class of the type A metric and the separability structure is indeed related to the
underlying generalised Killing--Yano symmetry.
As the Killing--Yano tensor by itself provides a constant of motion for the Dirac equation
\cite{Carter:1979,McLenaghan:1979,Kamran:1984,Benn:1997,Benn:2004,
Houri:2010b,Cariglia:2011,Cariglia:2011b,Carignano:2011},
it would be also interesting to study its separability in the presence of torsion.
In that case one might need to consider a deformed Dirac equation with torsion,
as discussed in \cite{Houri:2010,Houri:2010b}.
The separability might also be extended to other test field equations such as Maxwell's 
equations.
In our calculation, the torsion tensors associated with the generalised Killing--Yano symmetry
have been identified with the matter fields as $\bm{T} = \ast \bm{F} / \sqrt{3}$ in the
five-dimensional minimal supergravity and as $\bm{T} = \bm{H}$ in the abelian heterotic
supergravity, which is analogous to the asymptotically flat black hole spacetimes. 
In the limit of vanishing torsion, both of them reduce to the Kerr string solution.

The obtained classification provides an alternative to various approaches for finding
new exact solutions. In fact, using the type A metrics derived in Sec.~2, we have
constructed the general solutions describing charged, rotating Kaluza--Klein black
holes in the five-dimensional minimal supergravity and abelian heterotic supergravity.
Although we have concentrated on the type A metrics in this paper since our primary
aim has been the application to Kaluza--Klein black holes, type B and C metrics
also offer possibilities to find novel exact solutions, which can be interesting since
they exist only when the torsion is present. It should also be commented that all the
calculations in this paper can be generalised to odd dimensions higher than five. 
It would enable us to seek uplift of charged, rotating black hole solutions to
higher dimensions.

In \cite{Semmelmann:2002}, the existence of the ordinary Killing--Yano tensors 
was investigated on nearly K\"ahler manifolds and on manifolds with a weak $G_2$-structure.
The Killing--Yano equations on manifolds with G-structure were
investigated in the absence \cite{Papadopoulos:2008}
and presence \cite{Papadopoulos:2011} of torsion.
The generalised Killing--Yano tensors are also related to K\"ahler manifolds in even
dimensions and Sasaki manifolds in odd dimensions.
It was demonstrated in \cite{Houri:2012} that K\"ahler manifolds studied by \cite{Apostolov:2006} 
admit rank-2 generalised conformal Killing--Yano tensors. 
In odd dimensions, a concrete example
of the Killing--Yano tensor was constructed \cite{Visinescu:2012} on Sasaki manifolds 
studied by \cite{Gauntlett:2004}. 
Moreover, a notion of deformed Sasaki manifolds in the presence of torsion
was introduced by \cite{Houri:2012b} and the authors have shown an example 
admitting a rank-3 generalised closed conformal Killing--Yano tensor. 
It can be shown that the Sasaki manifolds with
torsion discussed in \cite{Houri:2012b} also take the form of the type A metrics 
in our classification. The present work might be useful to obtain further examples 
of Sasaki manifolds with torsion.

\ack
We are grateful to Gary W. Gibbons, David Kubiz\v{n}\'ak and Yukinori Yasui
for reading the manuscript and helpful comments.
T. H. would like to thank DAMTP, University of Cambridge, for the hospitality.
K. Y. would like to thank the Institute of Theoretical Astrophysics in the University
of Oslo for providing the stimulating environment.
The work of T. H. was supported by 
the JSPS Strategic Young Researcher Overseas Visits Program for
Accelerating Brain Circulation ``Deepening and Evolution of Mathematics and
Physics, Building of International Network Hub based on OCAMI.''

\end{document}